\documentclass[fleqn,twoside]{article}
\usepackage{espcrc2}
\readRCS $Id: espcrc2.tex,v 1.2 2004/02/24 11:22:11 spepping Exp $
\usepackage{graphicx}
\usepackage[figuresright]{rotating}
\newcommand{\be}{\begin{equation}}
\newcommand{\ee}{\end{equation}}
\newcommand{\ba}{\begin{array}{c}}
\newcommand{\ea}{\end{array}}
\newcommand{\bqa}{\begin{eqnarray}}
\newcommand{\eqa}{\end{eqnarray}}
\newcommand{\AmS}{{\protect\the\textfont2
  A\kern-.1667em\lower.5ex\hbox{M}\kern-.125emS}}
\title{Lightest scalars as chiral partners of the Nambu--Goldstone bosons}
\author{L.~Y.~Xiao\address[PKU]{Department of Physics,
        Peking University,
        Beijing 100871, China}%
        \,,
        H.~Q.~Zheng\addressmark\thanks{Invited talk given by Zheng at QCD06, Montpellier, France, July 3 -- 7, 2006. }
        \, and \,
        Z.~Y.~Zhou\addressmark\thanks{Present address: Department of Physics, Southeast University, Nanjing 210096, China}
        }

\begin{document}

\begin{abstract}
We review the spectrum of lightest scalar resonances recently
determined using dispersion techniques. The conceptual difference
between the pole mass and the bare mass (or the line--shape mass)
is stressed. The nature of the lightest scalars are discussed and
we argue, without relying on any model details, that the
$\sigma(500)$, $\kappa(700)$, $a_0(980)$ and $f_0(980)$ may be
understood as chiral partners of the Nambu--Goldstone bosons in
the linear realization of chiral symmetry. But there remains some
difficulties in understanding the role of $f_0(980)$ in this
picture.
\end{abstract}

% typeset front matter (including abstract)
\maketitle
\begin{figure}[htb]
\vspace{9pt} \framebox[75mm]{\rule[0mm]{0mm}{20mm}
\includegraphics[height=5.6cm]{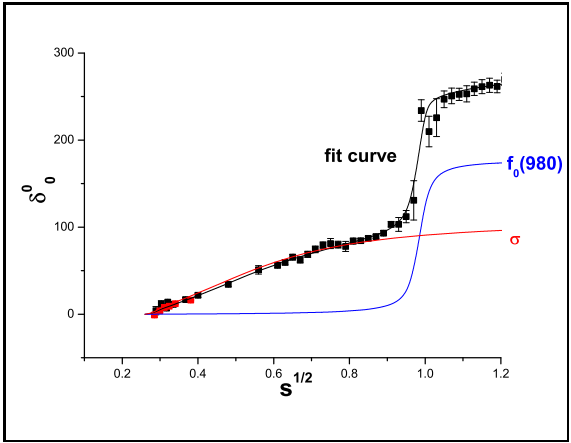}}
\caption{The $\sigma$ pole contribution to the IJ=00 channel
$\pi\pi$ scattering phase shift.   See Ref.~\cite{PKU2} for more
information. } \label{fig1}
\end{figure}

 The question on whether there exists a light and broad resonance with quantum number $IJ^P=00^+$
below 1GeV (for the historical reason, this resonance is properly
called the ``$\sigma$") had been a long standing issue of
debate~\cite{Tornqvist}. The major difficulties to accept $\sigma$
may be summarized as the following:
\begin{enumerate}
\item The $\sigma$ contribution in L$\sigma$M must exactly cancel the
$\pi\pi$ contact interactions, or more generally, background
contributions as dictated by PCAC and the soft pion theorem. In the
non-linear realization of chiral symmetry, the sigma meson is hence
not needed, at least in the very low energy region.
\item Being light and broad makes the $\sigma$'s contribution to the
phase shift hard to be distinguishable from background contributions
while the latter are often out of control in various
phenomenological studies.
\item The $\pi\pi$ phase shift in the I=0,
J=0 channel does not pass $\pi/2$ at moderately low energies which
should have appeared if there exist a light $\sigma$ as required by
the standard Breit--Wigner description of resonances.
\end{enumerate}
The above questions can be answered. In elastic $\pi\pi$
scatterings,  the background contribution, or the left hand cut
contribution can be estimated at low energies using chiral
perturbation theory. It is found that the background contribution to
the IJ=00 channel $\pi\pi$ scattering phase shift is negative and
concave while the experimental data is positive and convex, hence
demonstrating that the $\sigma$ meson is crucial to adjust chiral
perturbation theory to experiments~\cite{XZ00}, and hence the
existence of $\sigma$ is a must. Furthermore, in PKU parametrization
form~\cite{PKU1,PKU2,PKU3}, which may be considered as an upgraded
Dalitz--Tuan parametrization form, partial wave elastic scattering
amplitudes are constructed using dispersion technique, and the
amplitudes satisfy analyticity and single channel unitarity by only
assuming Mandelstam analyticity. In PKU form the pole contribution
to the phase shift is constructed explicitly and it is easy to
verify that, when there appears a large width, the pole mass
parameter, denoted as $m$, can be totally different from the mass
scale $M$ where the phase shift contributed by the pole pass
$\pi/2$~\cite{PKU1}. Taking the $\sigma$ resonance for example,
while $m\sim 450$MeV, $M$ is of order 1GeV. The two mass parameters
$m$ and $M$ looks quite different, but are actually only different
faces of a single light and broad resonance. The PKU form nicely
depicts and explains the strong enhancement of IJ=00 channel
$\pi\pi$ scattering phase shift over a long range of the center of
mass energy, which is named as ``red dragon" by Minkowski and
Ochs~\cite{MO99}. Fig.~\ref{fig1} depicts the situation, where one
sees clearly how a large width distorts the pole contribution to the
phase shift, comparing with the typical narrow resonance
$f_0(980)$'s contribution.
The background contribution is not depicted in fig.~\ref{fig1},
since its contribution is found to be small numerically. According
to this, it is possible to define the mass parameter $M$ for a pole
properly. Actually in the PKU parametrization form each pole's
contribution to the $S$ matrix is factorized:
\bqa\label{S}%
S^{pole}(s) = \frac{{M^2}(z_0)-s + i\rho(s)s G[z_0] }
  {{M^2}(z_0)-s - i\rho(s)s G[z_0]}\ ,
\eqa  where
 \bqa\label{mass}
&&{M^2}(z_0) = \mathrm{Re}[z_0] + \frac{\mathrm{Im}[z_0]\,
     \mathrm{Im}[z_0\,\rho (z_0)]}{\mathrm{Re}[
     z_0\,\rho (z_0)]}\ ,\nonumber\\
&&     G[z_0] =\frac{\mathrm{Im}[z_0]}{\mathrm{Re}[z_0\,\rho
(z_0)]}\ . \eqa
 In above the mass parameter $M$ is given as a function of the
pole mass ($\sqrt{z_0}\equiv m+i\Gamma/2$), $\rho$ is the kinematic
factor. The pole mass parameters of $\sigma$ and $\kappa$ are
determined to be (in units of MeV)~\cite{PKU2,PKU3}:
 \bqa\label{result pole}
&&m_\sigma=470\pm 50\ ,\,\,\,\,\,\, \Gamma_\sigma=570\pm 50\
,\nonumber\\
&&m_\kappa=694\pm 53\ ,\,\,\,\,\,\, \Gamma_\kappa=606\pm 59\ .
 \eqa
 These results, though with larger error bars, are fully compatible with the excellent results obtained using Roy
 equations: $m_\sigma=441^{+16}_{-8}$MeV, $\Gamma_\sigma=544^{+25}_{-8}$MeV~\cite{CCL}
 and $m_\kappa=658\pm 13$MeV, $\Gamma_\kappa=557\pm
 24$MeV~\cite{DM06}. Comparing with the sophisticated Roy equation analyses which demonstrate the existence
 of the sigma and kappa by direct calculation,
  the
 PKU approach is  more intuitive, simpler, and without much
 loss  of rigorousness (the theoretical uncertainties from the left hand cut in the latter
 approach is highly suppressed). It is worth emphasizing that, even if there is no
 Mandelstam representation, and hence analyticity of partial
 wave amplitudes is not established on the whole complex plane, it
 is proved over a large region~\cite{Martin} which is much
 larger than the analyticity domain straightforwardly established
 from Roy equations, and is certainly large enough for any phenomenological application.
 The PKU parametrization form is remarkable in the sense that it is  sensitive
 to the isolated singularities not too far from the elastic
 threshold. Except the determination on $\sigma$ and $\kappa$, it is also
 found to be useful in correctly figuring out the virtual state pole
 in the I=2 $s$ wave exotic channel~\cite{PKU2}, as well as the tiny effect of a virtual pole from the I=0
 $d$ wave $\pi\pi$ scattering phase shift data~\cite{PKU4}.
Comparing with the Roy equation analysis, the one adopted in the
present approach are orthogonal calculations, and it is important to
have  calculations  using different methods on such important
outputs and that the final results agree with each other. The value
on the pole positions of $\sigma$ and $\kappa$ obtained using
dispersion techniques, provide
 useful
 constraints when discussing the structure of the lightest QCD
scalars.

The first important information is the large width of $\sigma$ and
$\kappa$. The large widths are quite often overlooked when
discussing the mass spectrum in the literature. It is often
attempted to set up SU(3) mass relations among pole mass parameters
$m$. However, a light $\sigma$ with a mass around 500MeV as a bare
parameter appeared in the lagrangian can hardly produce a large
width, in any model calculations. On the other side,
 the parameter $M$ for $\sigma$ or $\kappa$ can be
determined using Eq.~(\ref{mass}) and the estimated value of the
pole positions are  $M_\sigma\simeq 930$MeV and $M_\kappa\simeq
1380$MeV with sizable error bars. It is noticed that the parameter
$M$ is very much like the bare mass parameter appeared in a simple
$K$ matrix approach, or the renormalized mass in the propagator of
loop calculations, since all of them take the role of determining
where the phase shift contributed by the resonating field passes
$\pi/2$. Parameter $M$ may also be called as the line--shape mass.
Qualitatively speaking the stronger the resonance couples to the
$\pi\pi$ continuum, the larger the deviation is between $m$ and $M$.
The situation is illustrated in fig.~2.
\begin{figure}[htb] \vspace{9pt}
\framebox[75mm]{\rule[0mm]{0mm}{20mm}
\includegraphics[height=5.4cm]{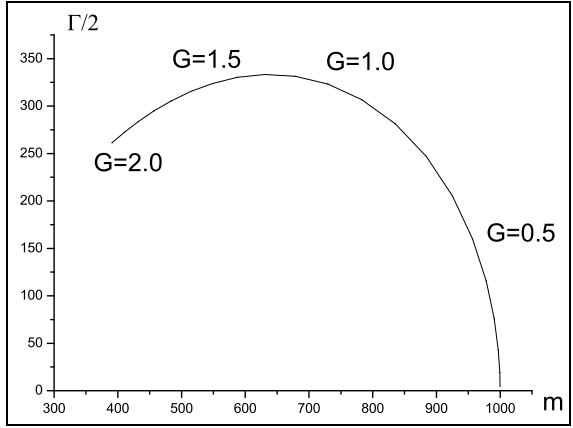}}
\caption{The pole trajectory determined from Eq.~(\ref{S}) as an
illustrative example. The bare mass is taken as 1GeV. The pole mass
decreases as the coupling strength, G, increases. The unit is MeV.}
\label{fig2}
\end{figure}
 If the comparison is correct, it will lead to an
important conceptual change when discussing the possible SU(3) mass
relations among lightest scalars: $\sigma$, $\kappa$, $a_0(980)$ and
also $f_0(980)$. Instead of comparing different $m$, on should first
examine  the relations between different ``bare" masses, $M$. Since
the former quantities associated with large widths are severely
distorted by the strong couplings to the pseudo-goldstone pairs, it
is not suitable to use them to discuss the SU(3) mass relations. For
example, we have
 \be\label{rel1}
m_\sigma<m_\kappa<m_{a_0}\ ,
 \ee
  but actually the mass relation should
be read as
 \be\label{rel2}
  1GeV\simeq M_\sigma  \leq M_{a_0}<M_\kappa\ .
 \ee
We will argue in the following that the relation Eq.~(\ref{rel2}) is
consistent with a chiral $\bar q q$ octet description to the
lightest scalars. Especially the bare mass of $\kappa$ exceeds
$M_{a_0}$, hence the original motivation leads to the tetra quark
description of the lightest scalars~\cite{Jaffescalar} already
disappear.  In fact it can be proved under mild assumptions, that a
tetra quark state (or more precisely, a tetra quark as the leading
component in Fock state expansions in the large $N_c$ limit) does
not exist since it violates analyticity~\cite{XZ05}. There exist a
few mass relations among the light scalars taking into account   't
Hooft's instanton effects, in the extended Nambu--Jona-Lasinio
model~\cite{su}:
 \bqa
&&M_{f_0}^2+M_{f_0'}^2\simeq 2M^2_{\kappa}-g_AM_{tH}^2\ ,\label{6}\\
&&M_\kappa^2-M_{a_0}^2\simeq 3g_A(M_K^2-M_\pi^2)\ ,\label{7} \eqa
 where $M_{tH}^2=M_\eta^2+M_{\eta'}^2-2M_K^2\simeq 0.72$GeV$^2$
characterizes
 the QCD anomaly contribution to the scalar
sector, and $g_A$ is the axial vector coupling due to $\pi$ -- $A_1$
mixing.
 The above
formulas are obtained by assuming that the scalars are chiral
partners of the pseudo-goldstone bosons. They are not exact
quantities but may serve as a crude ``first order" approximation.
 From the mass relations it is found that a
 smaller value of $M_\kappa$ and $g_A$ is preferred. Taking $g_A\simeq 0.6$ for example~\cite{Bijnens93}, if
 taking
 $M_\kappa=1.5$GeV, then through Eq.~(\ref{7}) one would attempt to
 consider the
 $a_0(1450)$ as a member of the chiral scalar octet. But then the
 well established conventional $\bar q q$ state $K^*_0(1430)$ will miss its own SU(3)
 partner,  which is not permitted. If   taking $M_\kappa=1.2$GeV,
 then Eq.~(\ref{7}) tells $M_{a_0}\simeq 1.02$GeV, i.e., not far from $a_0(980)$.
 The major difficulty in this picture comes from the mass of $f_0$.
  Eq.~(6) indicates $M_{f_0}\simeq 1.2$, if taking $M_{f_0'=\sigma}\simeq 1$GeV.
  Taking into account the  couple channel
 effects, which are large for $f_0(980)$, does not seem to be
 helpful in resolving the discrepancy.
Consider a simple Flatt\'e formula for a couple--channel
Breit--Wigner propagator,
 \be
BW(s)\propto \frac{1}{M^2-s-i\rho_1g_1^2-i\rho_2g_2^2}\ .
 \ee
 where the subscripts denotes different channels.
For a large $M^2$ above the second threshold this expression
generates a broad third sheet resonance and may also generate a
light resonance below the $\bar KK$ threshold. But unfortunately
the light one can only be located on the 4th sheet whereas the
observed $f_0(980)$ is on the second
 sheet.  For $f_0(980)$ the situation is  more
complicated since its bare state $f_0$ can also mix with
$f_0(1370)$, etc. Without instanton effects, the $\sigma$ and
$f_0$ are ideally mixed and the latter is $|\bar ss>$. When the
instanton effects are taken into account $f_0$ also contains
sizable $|\bar nn>$ content and hence may have a sizable mixing
with $f_0(1370)$, thus reducing significantly its mass. Hence we
think that the bare mass of $f_0(980)$ being roughly 250MeV above
the observed state is not a true difficulty in accepting the
picture that the $\sigma$, $\kappa$, $a_0(980)$ and $f_0(980)$
comprise the chiral partners of the pseudo-goldstone octet. This
idea is  not new~\cite{Tornq99}. The above discussions may however
be helpful in resolving some confusions and misconcepts widely
spread in the literature.
 One simple way to understand the difference
between the pole mass $m$ and the bare mass $M$  is through a
proper unitarization procedure~\cite{black01}. One advantage of
the chiral model with linearly realized chiral symmetry is that it
is not difficult to explain at qualitative level the large width
of $\sigma$ and especially the large width of $\kappa$. It is more
difficult to explain the latter~\cite{Narison05}, unless it
possesses a bare mass much larger than the pole mass parameter. At
lower energies one can then further integrate out the scalar
fields as well as other light resonances to get a  lagrangian with
non-linearly realized chiral symmetry.  We stress that the
``linear $\sigma$ model" here implies a more realistic linear
chiral model rather than the BPHZ renormalizable toy $\sigma$
model, since the latter failed to reproduce the LECs of the
Gasser--Leutwyler lagrangian~\cite{GL84}.

The use of dispersion techniques firmly establishes the existence
of $\sigma$ and $\kappa$, and also provides a reliable
determination to their mass parameters. The physical meaning of
these mass parameters are clarified and it is proposed that the
lightest scalars are chiral partners of the pseudo-goldstone
bosons. Nevertheless the role of $f_0(980)$ in such picture
remains to be clarified.

 {\bf Acknowledgement:} One of the author, H.Z., would like
to thank the organizer of QCD06, Stephan Narison, for providing a
charming atmosphere during the conference and helpful discussions,
and also Wolfgang Ochs, for interesting and helpful discussions in
the related subject. Especially H.Z. is grateful to Prof. San Fu
Tuan, for his kind encouragement. The authors would also like to
thank Zhi-Hui Guo and Ming-Xian Su for helpful discussions. This
work is supported in part by National Natural Science Foundation of
China under contract number
 10575002 and %(mianshang)
 10421503.%(tuandui)%and 10491306.%(BES)

\end{document}